# Life after delisting: tracking the publication output and citation impact of Scopus-discontinued journals with OpenAlex

Álvaro Cabezas-Clavijo & Fernando Sánchez-Pita

Universidad Internacional de La Rioja (UNIR)

Avenida de la Paz, 137, Logroño 26006, Spain

alvaro.cabezas@unir.net; fernando.sanchez@unir.net

*This manuscript is a preprint and has not undergone peer review. Comments, suggestions, and constructive feedback are welcome and appreciated to help improve the quality and clarity of the final version.*

**Abstract**

Selective databases such as Scopus periodically re-evaluate the titles they index and remove those that no longer meet their quality criteria. Once a journal is removed, its subsequent output is no longer recorded, and the consequences of this process have therefore received little attention. This study analyzes the publication output and citation impact of 514 sources delisted from Scopus between 2018 and 2024, based on 678,162 documents retrieved from OpenAlex over a window of three years before and after delisting. Differences between periods are tested with the Wilcoxon signed-rank test for paired samples, globally and by geographic region, field of knowledge and SJR quartile. Output grows until the year of delisting and falls sharply afterwards. The median annual output per journal drops from 35 to 22 documents (−37.1%; $p < 0.001$; $r = 0.30$), 61.7% of journals publish less, and 84 (16.3%) have no documents recorded in OpenAlex after delisting. Citations per document, measured over a three-year window, barely change when each journal is compared with itself (−2.4%; $p = 0.079$), and journals are almost evenly split between those whose impact falls and those whose impact holds steady or rises. The results indicate that delisting makes these journals less attractive as publication venues but does not penalize the citation of the work they publish to the same extent, which raises concerns from a research integrity perspective.

## Introduction

Delisted journals are publications that, having been admitted to a selective database, are removed from it years later because they no longer meet the quality criteria that justified their inclusion. This matters because journals, like people or brands, are valued for their reputation, trustworthiness and credibility.

The credibility of scientific activity rests on validation processes that operate at different levels of the system (Hicks et al., 2015; Saarela et al., 2026). For journals, peer review and editorial control act as guarantors of the reliability of what is published (Demir, 2018). For all their limitations, they help to curb the spread of false, low-quality or seriously flawed results (Elmore & Weston, 2020; Siler, 2020). This mechanism falls short, however, in an ecosystem that needs to classify and rank journals (Pölönen et al., 2021). Large databases such as Scopus and Web of Science (WoS) add a second layer of control. They act de facto as *whitelists* or trust markers, admitting titles only after expert committees have verified compliance with certain editorial standards (Chavarro et al., 2017; Singh et al., 2021). Besides providing visibility, indexing works as an implicit certificate of legitimacy (Bagues et al., 2019; Oviedo-García, 2021). As a result, these journal lists are routinely used in research evaluation in many countries (Marina & Sterligov, 2021).

This seal of quality also has a commercial value that attracts researchers. In a prestige-based academic economy, where publication is the main criterion for hiring and promotion (Trueblood et al., 2025), publishing in indexed or well-ranked journals earns credit that translates into promotion, direct financial incentives (Quan et al., 2017) or institutional funding (Mathies et al., 2020). This makes such journals preferred destinations for manuscripts.

This endorsement is not permanent, however. Selective databases periodically re-evaluate their titles and withdraw those that no longer meet their standards, a process known as delisting (Meho, 2025). Since withdrawal applies only prospectively, new content from the journal is no longer added to the database, while articles already indexed remain available and can still be found and cited (Cortegiani et al., 2020).

When a journal is excluded from one of the major scientific databases, a natural question is to what extent its standing among authors declines, that is, its capacity to attract new manuscripts, to sustain its editorial operations and to maintain its output as it did before delisting. It is equally reasonable to ask whether papers published after delisting achieve the same academic visibility and citation as those published earlier. This study addresses both questions through a bibliometric analysis that compares the output and impact of delisted journals before and after that point. To our knowledge, it is the first study of delisted journals to adopt this approach.

## Literature review

It is precisely the reputation attached to indexing that a range of questionable editorial practices exploit. Many of these practices have traditionally been grouped under the label of "predatory journals" (Beall, 2012). The consensus definition by Grudniewicz et al. (2019) characterizes these journals by their prioritization of self-interest over scholarship and by a set of behaviors such as false or misleading information, deviation from best editorial practices, lack of transparency, and aggressive and indiscriminate solicitation of authors.

These practices take many forms. Dadkhah et al. (2016) documented the hijacking of journal websites and phishing aimed at authors, while Cobey et al. (2018) compiled an extensive catalogue of traits attributed to predatory journals. Siler (2020), for his part, proposed distinguishing types and degrees of illegitimacy, a gradation that helps to interpret the already blurred boundary between the questionable and the fraudulent.

This supply is compounded by the pressure exerted by certain national evaluation systems that make career progression conditional on publishing in indexed journals. As a result, a considerable number of researchers end up taking part in questionable publishing in one way or another (Öztürk & Taşkın, 2024). This is one more instance of an almost Darwinian adaptation to the way academics are evaluated and rewarded (Butler, 2003; Jiménez-Contreras et al., 2003), regardless of any ethical or integrity considerations.

Nor does indexing make a journal immune to these dynamics. A title may qualify for inclusion in a database at one point and become "corrupted" later. The reasons vary. One is the pursuit of maximum profit, which under the pay-to-publish model encourages the acceptance of manuscripts that lack the required quality (Repiso-Caballero & Delgado-Vázquez, 2024). Another is the sale of the journal title. A network of companies that has acquired at least 36 journals indexed in WoS and Scopus since 2020 introduced or raised publication fees, increased output with papers from outside the journals' academic community and original scope, and even used fake DOIs (Martín-Martín & Delgado López-Cózar, 2025). The most serious cases point to organized networks in which paper mills, brokers selling authorships and compliant editors cooperate to publish papers that bypass peer review. In some instances, one of these networks has even taken full editorial control of a previously legitimate journal (Richardson et al., 2025).

The delisting of journals as a systematic practice is relatively recent. Selective databases already removed the occasional title that failed to meet their quality or impact criteria, but it was in the second decade of the twenty-first century that they began to act more forcefully against the contamination of the scholarly record by low-quality publications or those lacking adequate controls. In the case of Scopus, Nazarovets (2022) dates the start of the active re-evaluation of indexed titles to 2016. Scopus entrusts these periodic re-evaluations to its Content Selection and Advisory Board (CSAB), which applies to indexed titles the same criteria it requires of new ones (Table 1) and may discontinue coverage when a journal no longer meets them (Elsevier, 2025).

**Table 1. Scopus criteria for source selection.**

| Criterion | Requirements |
|---|---|
| Peer review | Publicly described review process. Type of review, timelines and acceptance rate. |
| Regularity and track record | Information on publication frequency, registered ISSN and established publishing history. Start year and annual number of articles. |
| Ethics and integrity | Public ethics statement and adherence to the 16 principles of transparency and best practice. Statement on generative AI, if any. |
| International reach | Titles and abstracts in English, geographic distribution of authors and indexing in other databases. |
| Editorial policy | Scope, publication languages, business model, open access and DOIs. |
| Editorial team | Public editorial board and profile of the editor-in-chief. |
| Identification and background | Unabbreviated title, subject area, publisher and country. Previous rejections by the CSAB. |

Source: Authors' elaboration based on Elsevier (2025).

The other major selective database, WoS, likewise warns that accepting a title does not guarantee permanent indexing, and subjects journals to periodic reviews that may end in the withdrawal of coverage (Quaderi, 2023)[1]. However, WoS did not begin to announce its delisted journals formally until 2023 (Meho, 2025).

In both cases, continued coverage depends on ongoing compliance with the criteria. Even so, not every exit from a database should be associated with malpractice. A title may disappear because it ceases publication, lacks quality submissions or loses institutional support (Jamali et al., 2022). It should also be noted that databases do not always agree on which journals should be delisted. Meho (2025) identified 548 journals delisted from Scopus and 184 from WoS between January 2018 and August 2025. Of the former, 82 were still indexed in WoS; of the latter, 71 remained in Scopus.

[1] Clarivate details the conditions for withdrawing journals on its page “Editorial policies, category change requests, and appeals process”. https://clarivate.com/academia-government/scientific-and-academic-research/research-discovery-and-referencing/web-of-science/web-of-science-core-collection/editorial-selection-process/editorial-policies-category-change-requests-appeals/

Sanctions, however, do not guarantee a change in behavior. In a study of 8,730 open access journals listed in DOAJ and indexed in Scopus, Castillo-Nuñez et al. (2026) found that a previous delisting was the strongest predictor of new episodes of instability. In addition, a substantial share of delisted journals continue to claim Scopus indexing after their removal (Krauskopf, 2018), thereby providing misleading information to the academic community.

The consequences of delisting have been studied from various angles. Cortegiani et al. (2020) analyzed the 317 journals that Scopus had discontinued for publication concerns up to mid-2019. Their articles, indexed before removal, continued to accumulate citations afterwards, and at a significantly higher annual rate than before. Lancho Barrantes et al. (2023) studied "citation contamination", a term attributed to Anderson (2019). It refers to the citation of articles published in questionable journals, a practice that threatens research integrity because it may help to spread potentially fraudulent work. Their analysis of 21 journals removed from DOAJ on suspicion of editorial misconduct identified 15,268 citations to these publications, mainly from Engineering, Computer Science and Mathematics. The authors caution, however, that citing one of these papers does not make the citing work questionable.

Castillo-Nuñez et al. (2026) examined the editorial factors associated with delisting. In their analysis, discontinuation was statistically related to the journals' history, financial transparency and governance capacity, and unrelated to routine editorial procedures such as plagiarism screening or response times. Journals that charged fees in addition to article processing charges (APCs) were almost four times more likely to be delisted. For these authors, delisting signals deeper structural vulnerabilities in the publishing system.

Another line of research comprises national studies that examine how these processes affect a country's output. Nazarovets (2022) analyzed publications by Ukrainian institutions in Scopus-delisted titles between 2011 and 2020. They accounted for 5.7% of national output, a lower share than in Romania (8.7%) or Russia (7.1%). Almost one in five Ukrainian authors had published at least one paper in them. Prevalence also varied across fields, peaking in Economics, Econometrics and Finance, where almost 47% of Ukrainian output had appeared in delisted journals.

Repiso-Caballero and Delgado-Vázquez (2024) studied the effect on the Spanish research system of the removal of 82 journals from WoS in 2023. Between 2018 and 2022, these titles had published 7,011 papers with Spanish authorship, 77.5% of them in a single journal, *International Journal of Environmental Research and Public Health*. In that journal alone, the estimated cost to Spanish authors exceeded 12 million euros over five years. Dependence was greater at small universities with lower research activity, and lower at those with clear research policies. The authors link this to an evaluation system that, far from penalizing these publications, rewards them as merits.

Öztürk and Taşkın (2024) documented a similar mechanism in Turkey, where publishing in questionable journals is associated with academic promotion processes. From a broader perspective, Marina and Sterligov (2021) measured the prevalence of potentially predatory journals in Scopus by country and linked the differences observed to the design of national evaluation policies.

These studies have focused on output indexed before delisting, because from the moment of removal the database stops recording what the journal publishes. Subsequent editorial activity has hardly been examined systematically, although there are signs that delisting reduces the attractiveness of these journals as publication venues (Richardson et al., 2025). It is at this intersection that the present study is situated.

## Objectives

The general objective is to analyze, from a bibliometric perspective, the changes associated with delisting from a major database such as Scopus in the journals subject to this exceptional measure.

The specific objectives are as follows:

- SO1. To determine the output of delisted journals before and after delisting, assessing whether there are significant differences globally, by field of knowledge, by the journal's geographic region and by its impact (SJR quartile) at the time of delisting.
- SO2. To analyze the citation impact of the journals, assessing whether there are significant differences in the citations received by papers published before and after delisting, according to the same variables as in SO1.

## Methodology

The study population comprised all sources delisted from Scopus between 2018 and 2024 (n = 563), according to the lists Scopus periodically publishes on its website[2], whose reason for removal was Discontinuation or Journal Change Policy. Although both labels are ambiguous, they usually cover problematic editorial behavior such as citation manipulation, artificial growth in publication volume or lack of rigor in peer review. Journals discontinued from Scopus for other reasons, such as simply ceasing publication, were not considered[3].

---

[2] https://www.elsevier.com/products/scopus/content

[3] For simplicity, the term “journals” is generally used throughout the text to refer to all the sources analyzed. According to the source types in the Scopus source list, however, 508 of the 514 sources

Using the Scopus master lists, we built a record of the journals that included, for each one, the country of publication, the country's geographic region (according to the regional classification used by the SCImago Journal & Country Rank) and its SJR quartile in the year of delisting. When no information was available for that year, the last available year before delisting was used. When a journal was ranked in more than one quartile in the same year, the most favorable quartile was recorded.

We also added the field of knowledge according to the six major areas of the Frascati Manual (OECD, 2015), in order to work at a level of aggregation that would make the results easier to interpret and compare. The starting point was the subject classification that Scopus assigns to each journal through the All Science Journal Classification (ASJC). Once the two classifications had been mapped, each journal was assigned to the area containing most of its categories. Finally, the reason for delisting was recorded as originally reported by Scopus.

Scopus could not be used to determine the output and citation impact of the sources before and after delisting, since it only contains information on papers published before that point. We therefore used OpenAlex, an open data source and an alternative to commercial databases (Priem et al., 2022), which is not affected by this limitation because it applies no selection filters to the journals it covers (Chavarro et al., 2025). Its coverage, which exceeds 250 million indexed publications, relies on a design based on persistent identifiers that facilitates interoperability and large-scale analysis (Forchino & Torres-Salinas, 2026). This openness has led to a more balanced representation of the social sciences and humanities and of regional output, as well as better coverage of diamond open access journals than Scopus or WoS offer (Simard et al., 2024; Céspedes et al., 2025). Its suitability for citation analysis in most fields has also been tested against Scopus (Thelwall & Jiang, 2025). This broader and more comprehensive coverage comes with limitations related to metadata quality and to the absence of quality controls on the sources it includes (Culbert et al., 2025).

For each of the 563 journals, we searched OpenAlex for output from the three years before delisting (t−3) to the three years after (t+3), where t is the year of delisting. Because the delisting cohorts extend to 2024 and output was considered up to 2025, output was recorded from t−3 to t+2 for journals delisted in 2023, and from t−3 to t+1 for those delisted in 2024.

Documents were collected with a script generated with Claude (Sonnet 4.6 model) that queried each journal's ISSN, which retrieved 530 journals (with or without associated output). After manual review, the script was modified to use the

---

are journals, five are conference proceedings series and one is a book series. These six sources represent 1.2% of the sample. They were retained in the analysis because they appear in the official list of discontinued titles with the reason Discontinuation.

OpenAlex source identifier instead of the ISSN, which recovered five additional journals. The scripts were run in Google Colab. For 21 of the 535 journals, a record existed in OpenAlex but no documents were retrieved, so these titles were excluded from the analysis.

The final sample therefore comprised the 514 delisted journals that were found in OpenAlex and had at least one document recorded in it. This represents 91.3% of the journals delisted from Scopus between 2018 and 2024. Over the period analyzed, these journals published a total of 678,162 papers. Table 2 shows the descriptive characteristics of the corpus.

**Table 2: Delisted sources and documents retrieved from OpenAlex by year of delisting**

| Year of delisting | Sources (N) | Sources in OpenAlex Documents>0 (N) | Coverage (%) | Docs (N) | Docs (%) |
|---|---|---|---|---|---|
| 2018 | 126 | 110 | 87.3 | 96,083 | 14.2 |
| 2019 | 38 | 38 | 100.0 | 55,505 | 8.2 |
| 2020 | 91 | 79 | 86.8 | 69,081 | 10.2 |
| 2021 | 80 | 73 | 91.3 | 205,440 | 30.3 |
| 2022 | 52 | 45 | 86.5 | 45,993 | 6.8 |
| 2023 | 83 | 81 | 97.6 | 106,405 | 15.7 |
| 2024 | 93 | 88 | 94.6 | 99,655 | 14.7 |
| **Total** | **563** | **514** | **91.3** | **678,162** | **100.0** |

Bibliographic records were then downloaded for all papers (678,162 documents) published by each of the 514 journals over a seven-year period (from t−3 to t+3). The exceptions are journals delisted in 2023, for which six years of data were retrieved (output from 2020 to 2025, that is, t−3 to t+2), and journals delisted in 2024, for which five years were retrieved (output from 2021 to 2025, that is, t−3 to t+1).

Second, a new script was written to collect the number of citations to the retrieved articles. To compare citations per paper (C/D) consistently regardless of the year of publication, we set a fixed three-year citation window covering the year of publication and the two following years. For each article, we therefore counted the citations received within this window rather than its total citations. For this reason, this indicator was analyzed only for papers published up to 2023, which had a sufficient citation window at the time of data collection (June 2026). As with output, citations were classified according to whether papers had been published before or after the journal's delisting.

*Statistical analysis*

To test differences in output and in citations per document before and after delisting, we applied the two-sided Wilcoxon signed-rank test for paired samples.

For each journal, the comparison was made between the median annual value in the pre-delisting period (t−3, t−2, t−1) and the median annual value in the post-delisting period (t+1, t+2, t+3). The year of delisting (t) was excluded from both periods because it contains papers published both before and after coverage was withdrawn. This test was chosen because of the skewed distribution of the output and citation variables.

The output analysis used the 514 journals in the valid corpus. For citations per document, the test was applied to the 276 journals with paired data, that is, with data available in both periods. This difference between the two corpora arises mainly because the citation analysis excludes output published after 2023, which leaves the cohorts delisted in 2023 and 2024 without an assessable post-delisting period, as they lack a complete citation window. The 84 journals with no output in the post-delisting period were also excluded from this test, since the citations per document indicator could not be computed for that period. In addition, 13 journals had data only in the post-delisting period and four only in the year of delisting, so they were also excluded from the Wilcoxon test.

The analysis was carried out for the corpus as a whole and stratified by geographic region, field of knowledge and SJR quartile. Effect size was estimated with the rank-biserial correlation (r), and the significance level was set at $\alpha = 0.05$.

Statistical analyses were performed in Python 3.13 with the scipy.stats library (Virtanen et al., 2020), and visualizations were produced in R 4.3.3 (R Core Team, 2024) with the ggplot2 package (Wickham, 2016).

## Results

### *Publication output*

Table 3 shows the journals analyzed by geographic region, field of knowledge and SJR quartile in the year of delisting (or the closest previous year available).

The corpus is clearly dominated by three regions that together account for more than three quarters of the delisted journals: Western Europe (30.4%), Asiatic Region (26.8%) and Northern America (19.1%). Journals published in Latin America (n = 21) have the highest median number of articles per journal (833), in contrast to the other regions, whose medians lie between 300 and 400 articles per journal. Only Africa (n = 6) departs from this pattern, with a median of 251.5 papers per journal and 1.1% of the documents in the sample.

The three most represented fields of knowledge are Medical and Health Sciences (25.9%), Engineering and Technology (24.3%) and Social Sciences (23.7%), which together account for almost three quarters of the corpus. Engineering and Technology contributes the largest number of documents to the analysis (more than 300,000, or 44.5% of the sample).

Most of the delisted sources were in the fourth (44.9%) or third (28.8%) SJR quartile, with a minority in the second (15.8%) or first quartile (5.4%).

**Table 3: Production of delisted sources by geographic region, field of knowledge and SJR quartile.**

| | Sources (N) | Sources (%) | Docs (N) | Docs (%) | Median | Mean |
|---|---|---|---|---|---|---|
| ***Region*** | | | | | | |
| Western Europe | 156 | 30.4 | 283,907 | 41.9 | 325.0 | 1,819.9 |
| Asiatic Region | 138 | 26.8 | 172,571 | 25.4 | 378.5 | 1,250.5 |
| Northern America | 98 | 19.1 | 101,440 | 15.0 | 358.0 | 1,035.1 |
| Eastern Europe | 46 | 8.9 | 35,521 | 5.2 | 371.0 | 772.2 |
| Middle East | 36 | 7.0 | 39,007 | 5.8 | 399.0 | 1,083.5 |
| Latin America | 21 | 4.1 | 29,462 | 4.3 | 833.0 | 1,403.0 |
| Pacific Region | 13 | 2.5 | 8,662 | 1.3 | 356.0 | 666.3 |
| Africa | 6 | 1.2 | 7,592 | 1.1 | 251.5 | 1,265.3 |
| ***Field of Knowledge*** | | | | | | |
| Medical and Health Sciences | 133 | 25.9 | 164,936 | 24.3 | 387.0 | 1,240.1 |
| Engineering and Technology | 125 | 24.3 | 301,971 | 44.5 | 435.0 | 2,415.8 |
| Social Sciences | 122 | 23.7 | 86,975 | 12.8 | 357.0 | 712.9 |
| Natural Sciences | 82 | 16.0 | 95,646 | 14.1 | 366.0 | 1,166.4 |
| Humanities and the Arts | 27 | 5.3 | 15,882 | 2.3 | 317.0 | 588.2 |
| Agricultural Sciences | 25 | 4.9 | 12,752 | 1.9 | 244.0 | 510.1 |
| ***SJR Quartile*** | | | | | | |
| Q1 | 28 | 5.4 | 21,115 | 3.1 | 200.0 | 754.1 |
| Q2 | 81 | 15.8 | 119,603 | 17.6 | 612.0 | 1,476.6 |
| Q3 | 148 | 28.8 | 156,474 | 23.1 | 465.0 | 1,057.3 |
| Q4 | 231 | 44.9 | 242,527 | 35.8 | 273.0 | 1,049.9 |
| No SJR quartile | 26 | 5.1 | 138,443 | 20.4 | 479.5 | 5,324.7 |
| **Total** | **514** | **100.0** | **678,162** | **100.0** | **365.0** | **1,319.4** |

Table 4 shows how the different statistics evolve in the years before and after delisting. Annual output fell markedly after delisting. Journals published a median of 35 documents per year before delisting and 22 afterwards, a decrease of 37.1%. The Wilcoxon signed-rank test confirms that the difference is statistically significant ($p < 0.001$), with a moderate effect size ($r = 0.30$). For the 2023 and 2024 cohorts, the post-delisting period includes only the years observed up to 2025. When only the 2018–2022 cohorts, which have all three post-delisting years, are considered, median output falls from 30 documents before delisting to 17 afterwards, a decrease of 43.3% ($p < 0.001$; $r = 0.36$).

At the journal level, 61.7% of journals reduced their output after delisting. Indeed, 84 journals, 16.3% of the sample, have no papers recorded in OpenAlex after delisting. The decline is not driven solely by journals that stop publishing, however. When journals with no documents in either the pre- or the post-delisting period are excluded, the median annual output per journal falls from 43 to 31 documents (−27.9%; $p < 0.001$; $r = 0.24$), with 60.3% of journals reducing their output.

**Table 4: Production of delisted sources before and after delisting**

| Time | N sources | N docs | Median | Mean |
|---|---|---|---|---|
| t−3 | 514 | 74,132 | 29.0 | 144.2 |
| t−2 | 514 | 95,457 | 32.0 | 185.7 |
| t−1 | 514 | 141,640 | 49.5 | 275.6 |
| ***Pre-delisting period*** | ***514*** | ***311,229*** | ***35.0*** | ***201.8*** |
| t | 514 | 200,224 | 64.0 | 389.5 |
| t+1 | 514 | 108,766 | 30.0 | 211.6 |
| t+2 | 426 | 36,555 | 16.0 | 85.8 |
| t+3 | 345 | 21,388 | 11.0 | 62.0 |
| ***Post-delisting period*** | ***514*** | ***166,709*** | ***22.0*** | ***150.4*** |

Figure 1 shows the growth in output up to the year of delisting and the sharp fall after the journals leave Scopus. Median output rises from 29 documents in t−3 to 49.5 in t−1 and peaks in the year of delisting (64 documents). In just two years, between t−2 and t, median output doubles (from 32 to 64 documents). After removal from Scopus, output drops sharply. The median falls to 30 documents in the following year (−53.1% relative to year t), a level similar to that of the years before the peak, and continues to decline to 16 documents two years later and 11 in the third year. Overall, 67.5% of journals published less in t+1 than in the year of delisting.

**Figure 1: Yearly documents per source before and after Scopus delisting**

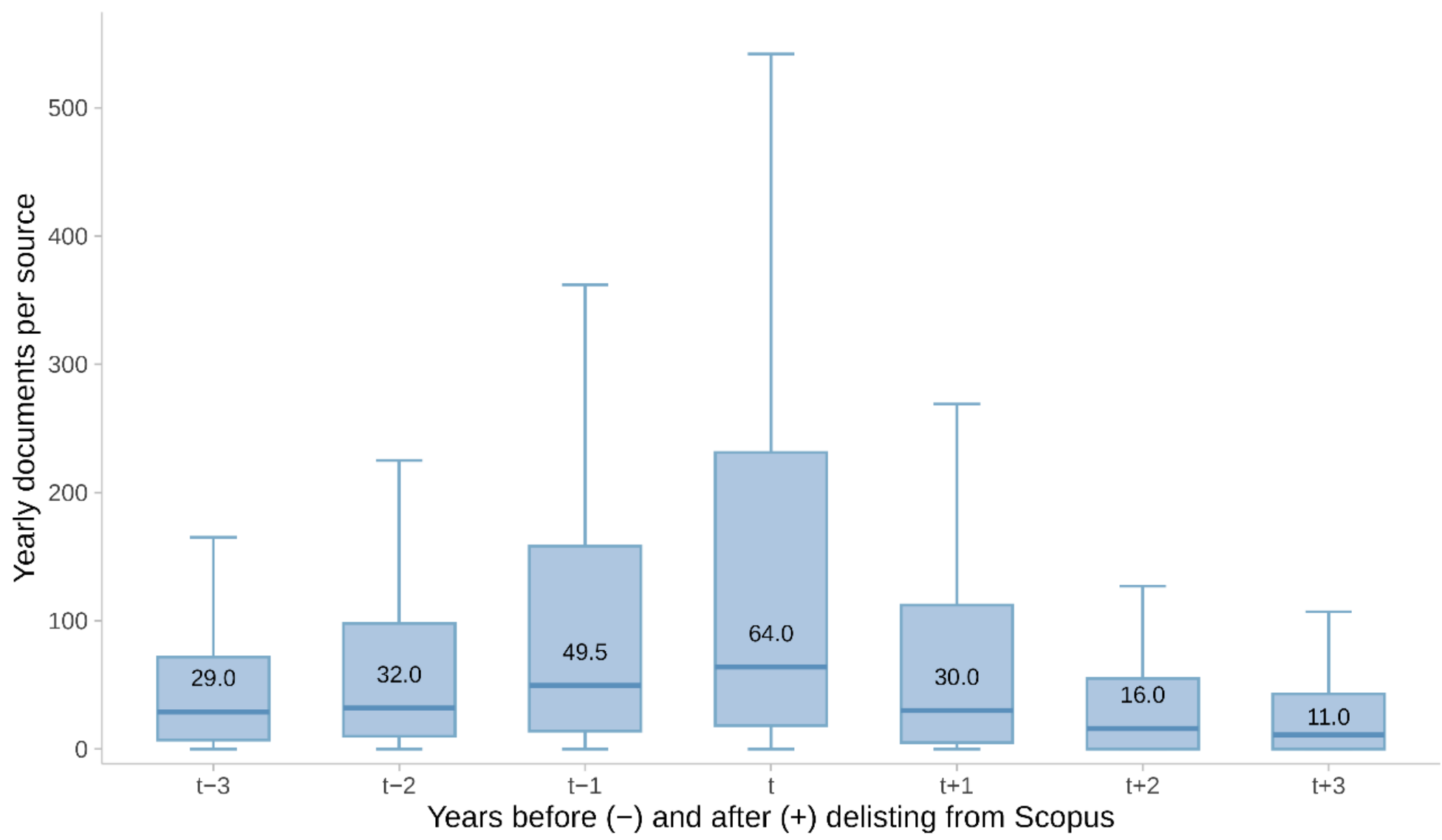

The general pattern of growth up to the year of removal followed by a decline is repeated in every geographic region (Figure 2), although the size of the decline and its statistical significance vary considerably. The largest falls in median output are found in Pacific Region (−87.8%), Middle East (−83.5%), Latin America (−54.7%) and Northern America (−54.1%). The decline is statistically significant in four regions: Northern America ($p < 0.001$; $r = 0.394$), Western Europe ($p < 0.001$; $r = 0.324$), Eastern Europe ($p = 0.036$; $r = 0.364$) and Asiatic Region ($p = 0.007$; $r = 0.275$).

In every region, between 60% and 65% of journals reduce their output after delisting. The exceptions are Middle East, where 55.6% of journals reduce their activity, and Africa, where only two of the six journals show a decline.

**Figure 2: Yearly documents per source before and after Scopus delisting by geographic region**

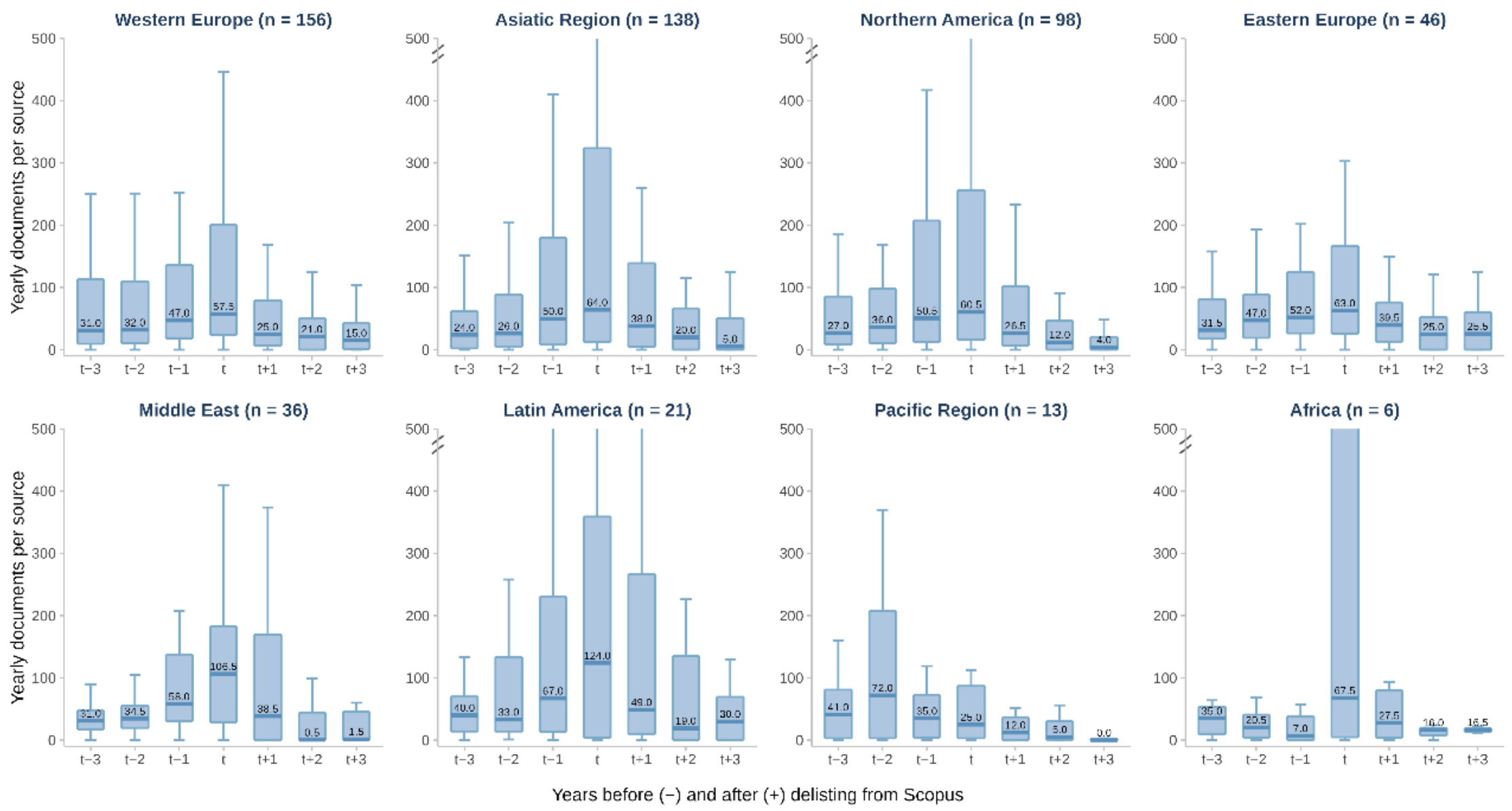

The same pattern appears in all fields of knowledge (Figure 3): output grows steadily until the year of delisting and declines in the following years. Engineering and Technology shows the steepest fall in median output (−53.5%; $p < 0.001$; $r = 0.437$), followed by Medical and Health Sciences (−32.5%; $p = 0.007$; $r = 0.278$) and Social Sciences (−31.3%; $p = 0.001$; $r = 0.343$), all of them statistically significant. In the other three fields the differences are not significant.

Even so, in every field more journals reduce their output after delisting than increase it. The extreme values are found in Agricultural Sciences, where 72% of journals reduce their output, and Natural Sciences, which, at 51.2%, is the field with the smallest share of journals in decline. In the remaining fields, between 55% and 64% of journals reduce their output.

**Figure 3: Yearly documents per source before and after Scopus delisting by field of knowledge**

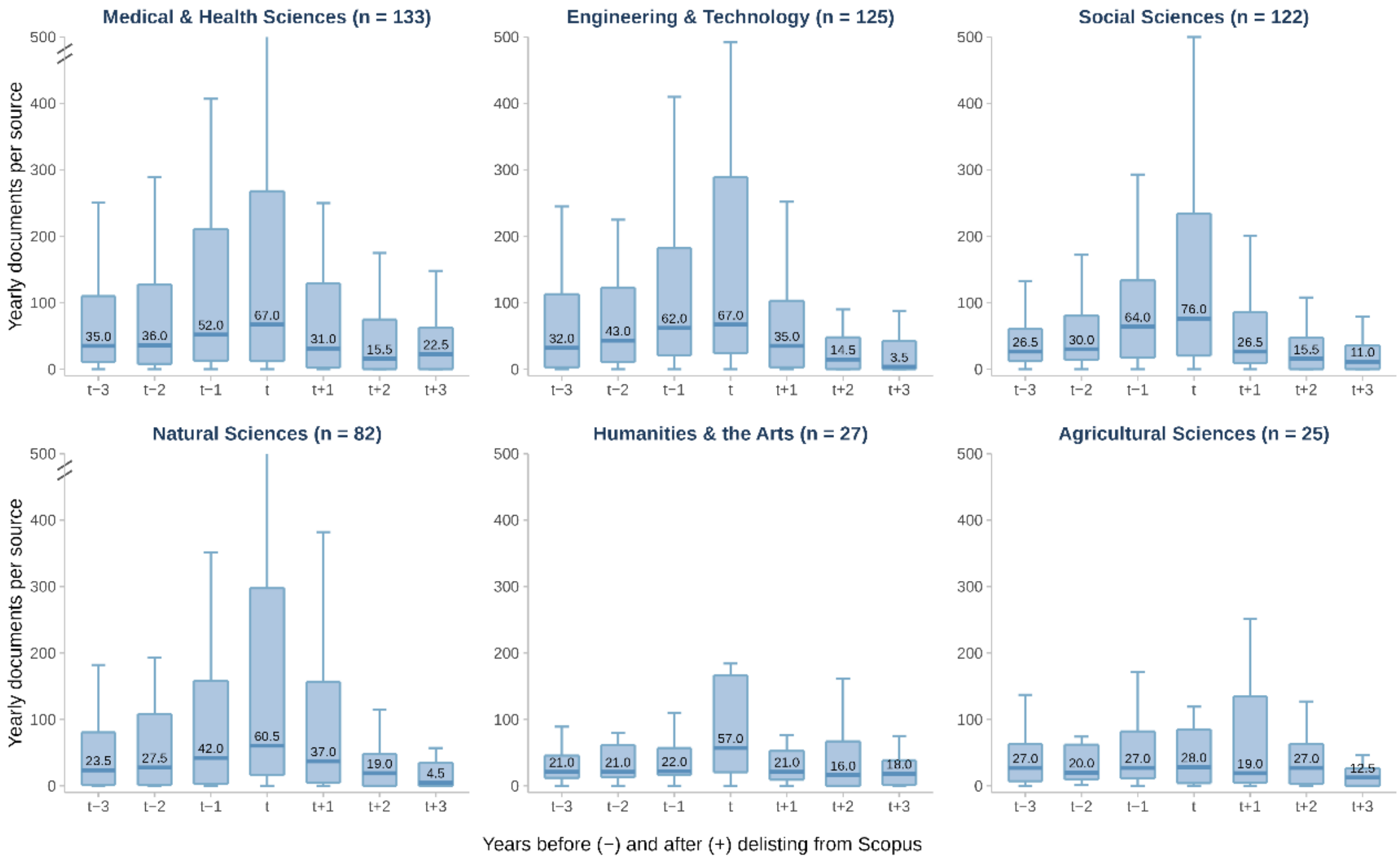

The general pattern of growth up to the year of delisting followed by a decline also holds across SJR quartiles (Figure 4), although the size of the effect differs. Q2 journals show the steepest decline (−45.6% in the median) and the largest effect size ($p < 0.001$; $r = 0.505$), with 67.9% of journals reducing their output after delisting. They are followed by Q4 (−44.0%; $p < 0.001$; $r = 0.288$) and Q3 journals (−41.9%; $p = 0.002$; $r = 0.304$), with more than 60% of journals reducing their output in both groups. In Q1 journals, by contrast, the change is not statistically significant. Their median output rises slightly (from 20 to 21.5 documents; +7.5%; $p = 0.269$), although 60.7% of these journals reduce their output relative to the previous period.

**Figure 4: Yearly documents per source before and after Scopus delisting by SJR quartile**

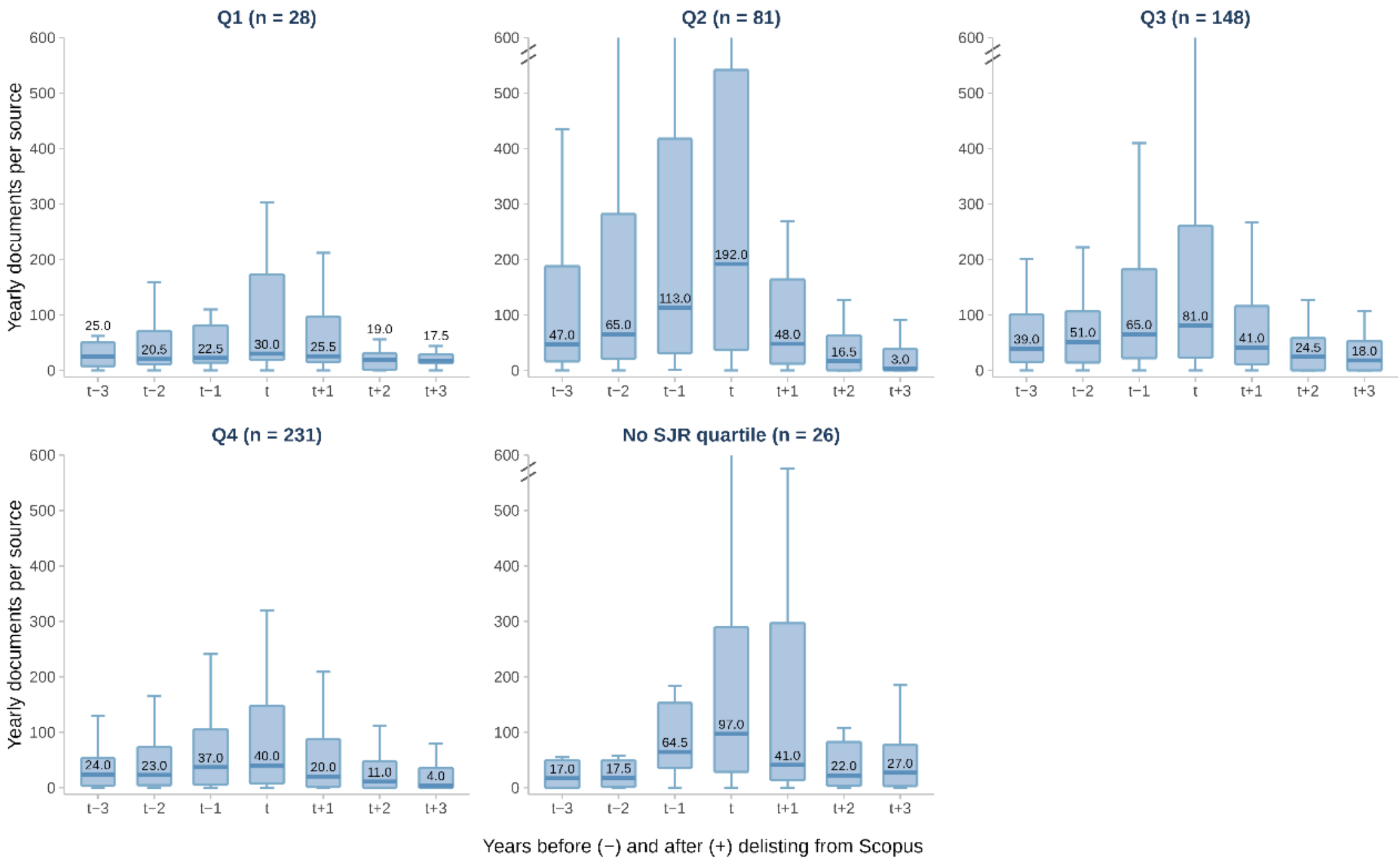

*Citations*

The citation analysis is based on the 581,540 documents published up to 2023 by 509 journals, since papers published in 2024 and 2025 do not have a complete three-year citation window. For the sample as a whole (Table 5), the median is 1.1 citations per document (mean 2.4). Almost half of the documents (49.3%) receive no citations within this window.

By geographic region, Northern America, Middle East and Pacific Region have the highest medians (1.6 citations per document). Northern America also has the highest share of cited documents (60.5%), followed by Western Europe (55.7%). By field of knowledge, Social Sciences has the highest median citations per document (1.6), followed by Engineering and Technology (1.2) and Natural Sciences (1.1), which also has the highest share of cited documents (58.8%).

By quartile, Q2 journals have the highest median citations per document (2.5), followed by Q1 journals (2.0), although the latter have a higher mean (5.4 versus 3.9). The share of documents receiving at least one citation is directly related to the journal's quartile, ranging from 71.9% in Q1 to 43.1% in Q4, which also has the lowest median citations per document (0.7).

**Table 5: Citation impact of delisted sources by geographic region, field of knowledge and SJR quartile**

| | Sources (N) | Docs (N) | Docs (%) | Cited docs (%) | Median C/D | Mean C/D |
|---|---|---|---|---|---|---|
| ***Region*** | | | | | | |
| Western Europe | 154 | 253,520 | 43.6 | 55.7 | 1.0 | 2.0 |
| Asiatic Region | 138 | 157,022 | 27.0 | 40.7 | 1.0 | 1.5 |
| Northern America | 95 | 81,639 | 14.0 | 60.5 | 1.6 | 3.0 |
| Eastern Europe | 46 | 25,548 | 4.4 | 47.8 | 1.0 | 1.7 |
| Middle East | 36 | 33,829 | 5.8 | 39.7 | 1.6 | 2.6 |
| Latin America | 21 | 22,198 | 3.8 | 48.2 | 1.2 | 4.2 |
| Pacific Region | 13 | 7,002 | 1.2 | 48.9 | 1.6 | 11.5 |
| Africa | 6 | 782 | 0.1 | 50.0 | 1.5 | 1.6 |
| ***Field of Knowledge*** | | | | | | |
| Medical and Health Sciences | 132 | 135,450 | 23.3 | 47.6 | 0.7 | 1.9 |
| Engineering and Technology | 125 | 276,152 | 47.5 | 50.2 | 1.2 | 3.1 |
| Social Sciences | 121 | 67,462 | 11.6 | 52.7 | 1.6 | 2.9 |
| Natural Sciences | 79 | 79,429 | 13.7 | 58.8 | 1.1 | 1.8 |
| Humanities and the Arts | 27 | 12,332 | 2.1 | 32.0 | 0.9 | 1.7 |
| Agricultural Sciences | 25 | 10,715 | 1.8 | 48.4 | 1.0 | 1.5 |
| ***SJR Quartile*** | | | | | | |
| Q1 | 28 | 15,851 | 2.7 | 71.9 | 2.0 | 5.4 |
| Q2 | 81 | 99,360 | 17.1 | 67.6 | 2.5 | 3.9 |
| Q3 | 148 | 118,841 | 20.4 | 53.8 | 1.4 | 2.1 |
| Q4 | 226 | 214,644 | 36.9 | 43.1 | 0.7 | 1.8 |
| No SJR quartile | 26 | 132,844 | 22.8 | 44.9 | 0.9 | 1.4 |
| **Total** | **509** | **581,540** | **100.0** | **50.7** | **1.1** | **2.4** |

*C/D = Citations / Document*

Table 6 shows the evolution of impact per document in the years before and after delisting. Median citations per document remain between 1.1 and 1.3 in the years before delisting, fall slightly in the year of delisting (1.2) and continue to decline in the following years, down to 0.7 in t+3 (Figure 5). For the period as a whole, the median falls from 1.2 citations per document before delisting to 0.9 afterwards.

These results should be interpreted with caution. The analysis is restricted to papers with a complete citation window, which progressively reduces the sample in the years after delisting. The post-delisting period comprises 103,539 documents from 289 journals, roughly one third of those in the pre-delisting period (311,229 documents from 492 journals), and the value for t+3 is based on only 12,120 documents from 154 journals.

**Table 6: Citation impact of delisted sources before and after delisting**

| Time | Cohorts | N sources | N docs | Cited docs (%) | Median C/D | Mean C/D |
|---|---|---|---|---|---|---|
| t−3 | 2018–2024 | 445 | 74,132 | 55.1 | 1.1 | 2.1 |
| t−2 | 2018–2024 | 451 | 95,457 | 56.1 | 1.3 | 2.4 |
| t−1 | 2018–2024 | 464 | 141,640 | 56.4 | 1.3 | 2.9 |
| ***Pre-delisting period*** | ***2018–2024*** | ***492*** | ***311,229*** | ***56.0*** | ***1.2*** | ***2.5*** |
| t | 2018–2023 | 381 | 166,772 | 46.4 | 1.2 | 2.1 |
| t+1 | 2018–2022 | 282 | 68,482 | 39.4 | 1.1 | 3.7 |
| t+2 | 2018–2021 | 209 | 22,937 | 46.3 | 0.9 | 1.4 |
| t+3 | 2018–2020 | 154 | 12,120 | 44.1 | 0.7 | 1.4 |
| ***Post-delisting period*** | ***2018–2022*** | ***289*** | ***103,539*** | ***41.5*** | ***0.9*** | ***2.4*** |

*C/D = Citations / Document*

Moreover, the paired comparison between periods for the 276 journals with citation data before and after delisting shows no statistically significant differences ($p = 0.079$; $r = 0.124$). For these journals, median citations per document barely change (from 0.93 to 0.90; −2.4%), compared with the 28.2% decline in the pooled median (from 1.2 to 0.9). At the journal level, 51.1% of journals reduce their C/D after delisting, while 48.9% maintain or even increase it. This near-even split explains the absence of a significant effect in the test.

**Figure 5: Yearly citations per document before and after Scopus delisting**

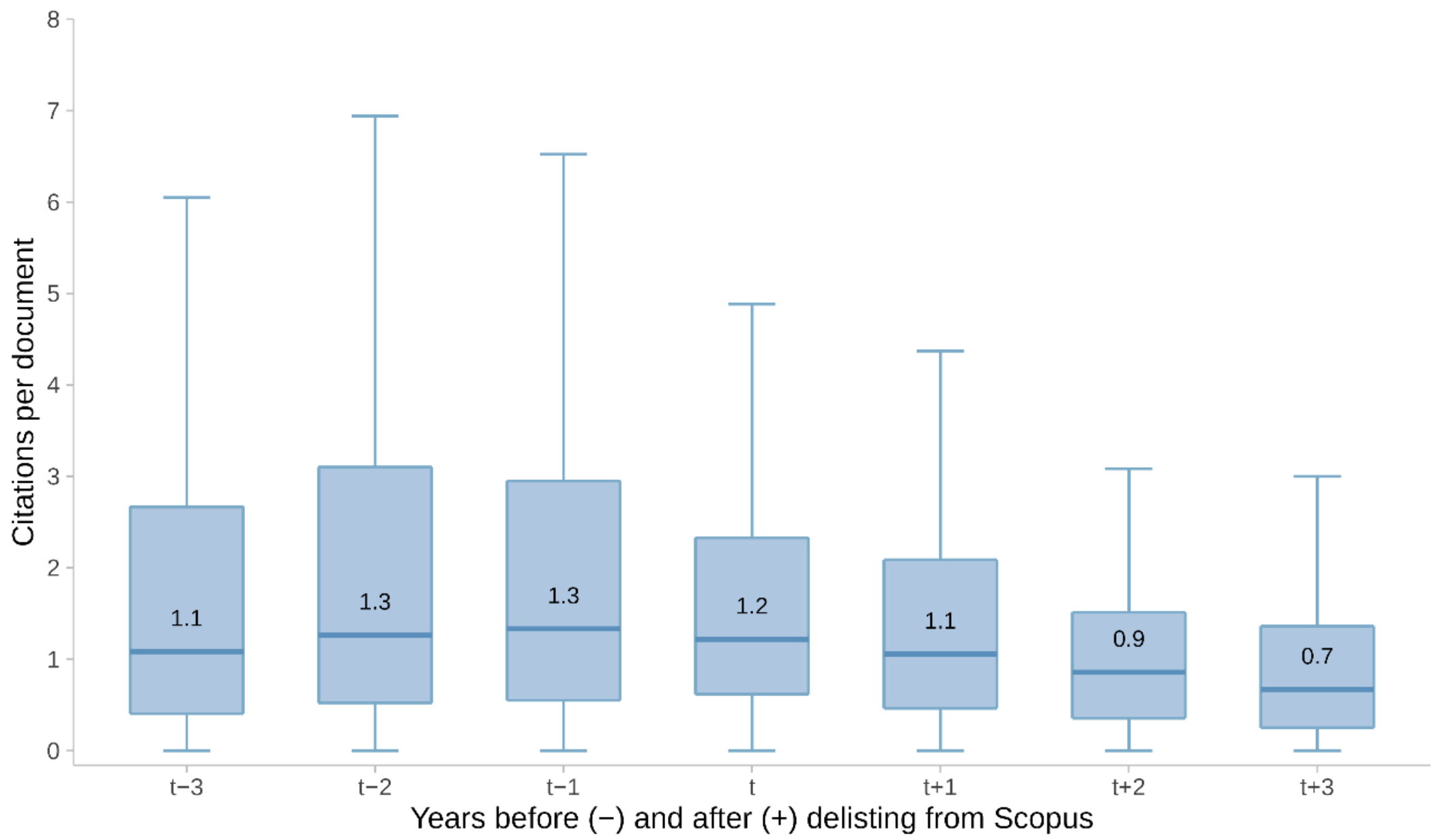

Across the eight geographic regions (Figure 6), median citations per document fall in five and rise in the other three. The steepest declines between the pre- and post-delisting periods are in Northern America (from 1.9 to 1.0; −46.4%), Western Europe (from 1.1 to 0.6; −44.4%) and Middle East (from 1.8 to 1.1; −39.4%). Latin America (from 1.0 to 1.4; +31.4%), Pacific Region (from 1.3 to 3.0; +135.1%) and Africa (from 0.9 to 2.5; +184.0%) show increases, although their samples are small.

The Wilcoxon test detects significant differences only in Middle East, based on 18 journals (p = 0.034; r = 0.567), and in Northern America, with 49 journals (p = 0.041; r = 0.347). In the former, 72% of journals reduce their C/D after removal from Scopus, compared with 65% in Northern America.

In Western Europe (77 journals analyzed), 53% of journals maintain or increase their C/D (p = 0.777; r = 0.038), and in Eastern Europe (28 journals) the figure is 57% (p = 0.737; r = 0.074). Asiatic Region (86 journals) shows a similar picture, with an even split between journals whose impact rises and those whose impact falls (p = 0.063; r = 0.231).

**Figure 6: Yearly citations per document before and after Scopus delisting by geographic region**

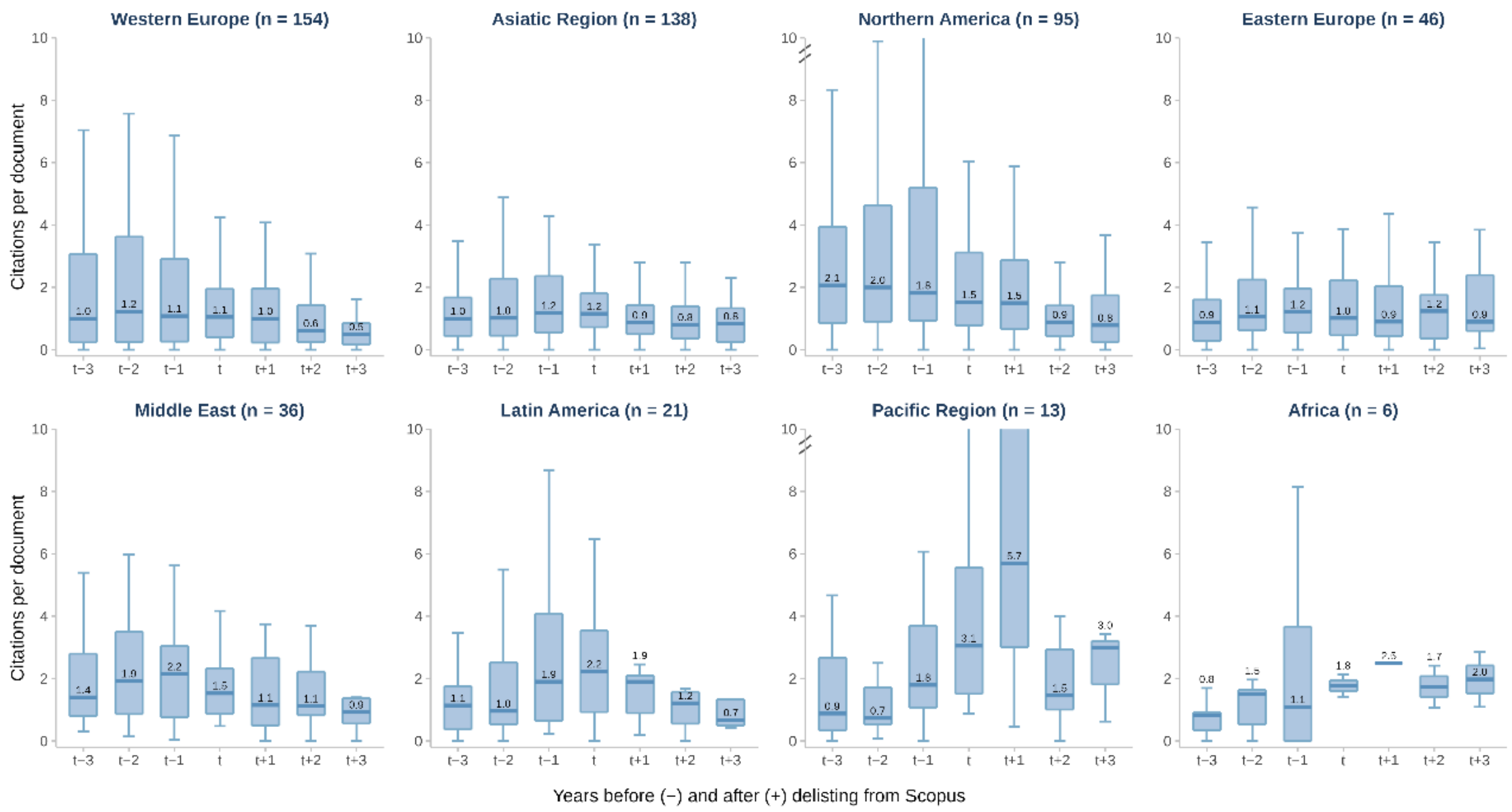

By field of knowledge (Figure 7), Medical and Health Sciences shows the steepest fall in median citations (from 0.9 to 0.4; −51.7%), although journals are almost evenly split between those whose C/D falls and those whose C/D holds steady or rises (52% versus 48%, out of 67 journals with data in both periods; p = 0.058; r = 0.272). Social Sciences has the highest share of journals with a lower C/D (59%) and a 28.8% fall in median citations (from 1.8 to 1.3), bordering on statistical significance (p = 0.050; r = 0.286).

In Engineering and Technology, although the median falls from 1.3 citations in the pre-delisting period to 1.0 in the post-delisting period, 56% of the 72 journals with data in both periods maintain or increase their C/D after delisting (p = 0.463; r = 0.100). Natural Sciences shows a similar pattern, with a 26.9% fall in median citations (from 1.3 to 0.9) and 55% of journals declining on this indicator (p = 0.472; r = 0.130).

Humanities and the Arts is the only field in which median citations are higher in the post-delisting period, rising from 0.8 to 0.9 (p = 0.934; r = 0.033). At the journal level, however, 53% of journals reduce their C/D. Finally, in Agricultural Sciences, despite the fall in median citations from 1.1 to 0.9, 69% of the 16 journals with paired data maintain or increase their C/D after delisting (p = 0.528; r = 0.191).

**Figure 7: Yearly citations per document before and after Scopus delisting by field of knowledge**

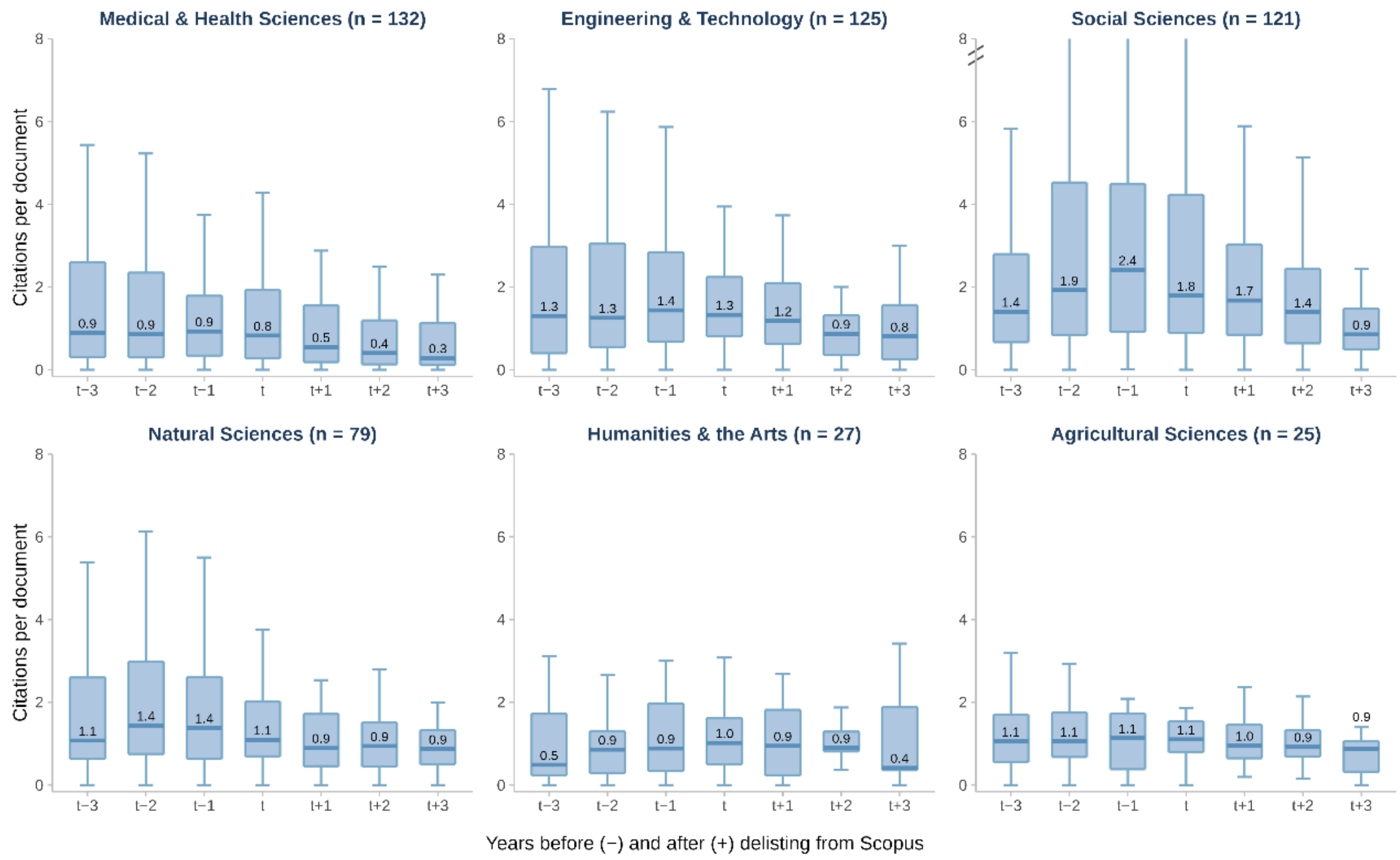

By journal quartile (Figure 8), median citations fall in all four quartiles, though to different degrees. Q1 journals, the smallest group (28 journals, of which only 14 have paired data), show a decline of 29.1% (from 2.6 to 1.8 citations per document), with 64% of journals reducing their C/D after delisting, and the difference is statistically significant (p = 0.033; r = 0.670). Q2 journals (81, of which only 36 have paired data) show the largest fall in citations (−55.0%, from 3.5 to 1.6) and the same share of declining journals as Q1 (64%), but the difference does not reach statistical significance (p = 0.132; r = 0.292).

In Q3 journals (148 journals, 82 with data in both periods), median citations fall from 1.6 to 1.1, although 51% of these journals maintain or increase their C/D (p = 0.313; r = 0.131). The same holds for Q4 journals (226 journals, 127 with paired data). Despite a slight fall in median citations (from 0.7 to 0.6), 57% of them maintain or even increase their C/D after delisting (p = 0.302; r = 0.106).

**Figure 8: Yearly citations per document before and after Scopus delisting by SJR quartile**

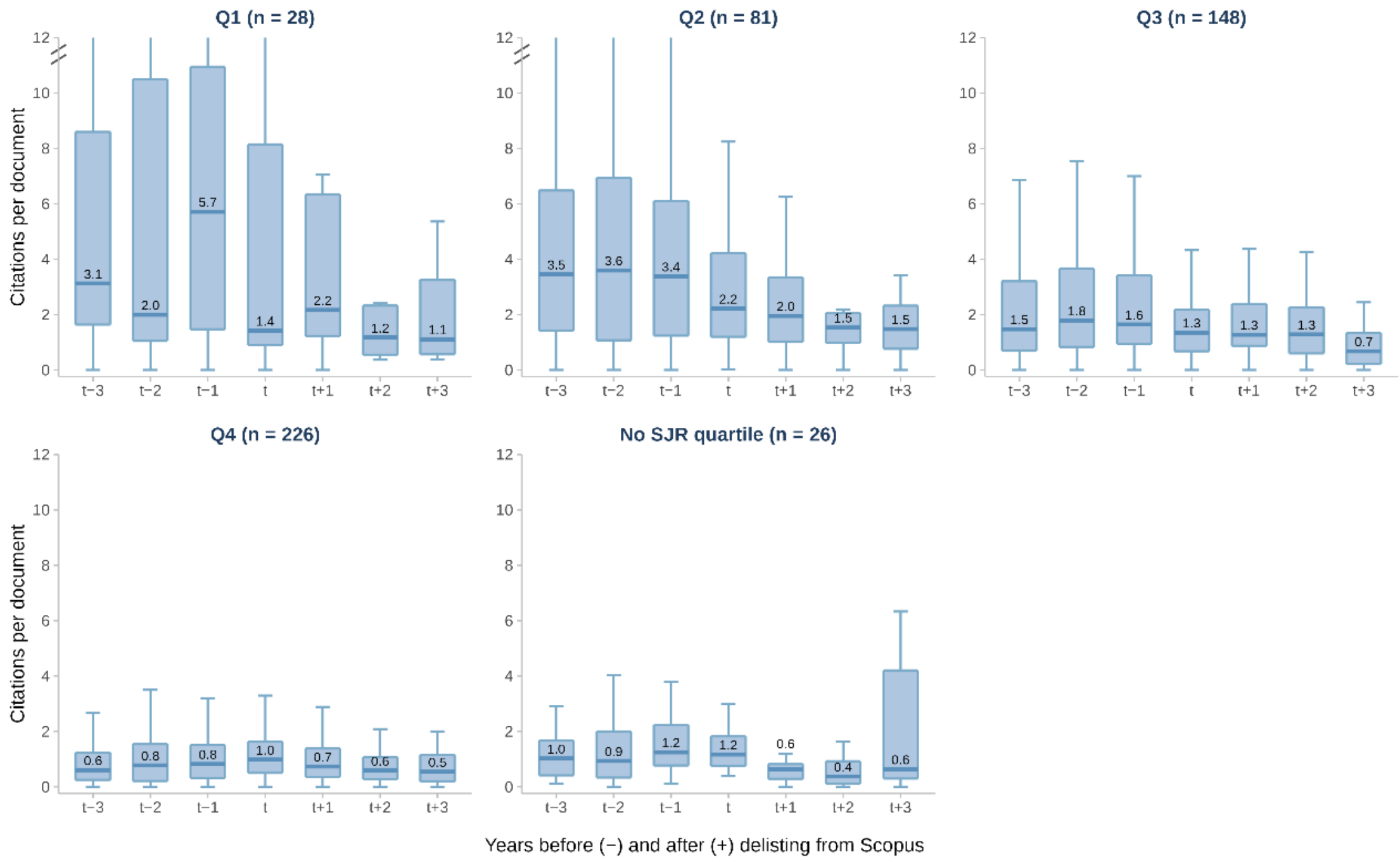

## Discussion

This study documents, for the first time, the output and citation patterns of sources delisted from Scopus between 2018 and 2024, with the aim of determining the changes in scientific output and citation associated with a journal's delisting.

One of the most relevant findings is that delisting affects output and impact per document differently. After removal from Scopus, median output per journal falls by 37.1%, a statistically significant decline of moderate size (r = 0.30) that affects six out of ten journals (61.7%). Citations per document, by contrast, barely change when each journal is compared with itself (−2.4%), and journals are almost evenly divided between those whose impact falls (51.1%) and those whose impact holds steady or rises, with no statistically significant differences.

Output per journal falls clearly, from a median of 35 papers per year before delisting to 22 afterwards. Output grows steadily up to the year of delisting, after which activity drops sharply. In addition, 84 journals (16.3%) have no papers recorded in OpenAlex after delisting. The decline is observed, to a greater or lesser extent, in all geographic regions and fields of knowledge, although it is statistically significant only in four of the eight regions (led by Northern America) and in three of the six fields, with the steepest fall in Engineering and Technology (−53.5%).

A further analysis of these 84 journals found that slightly more than half (54%) have in fact ceased publication, whereas the rest continue to publish and maintain an active website but are not covered by OpenAlex. Several reasons may explain why some delisted journals keep publishing without their output being recorded in this database. OpenAlex draws on multiple sources but relies heavily on the metadata that publishers deposit with Crossref when registering DOIs. If a journal stops doing so, whether for financial reasons or because it uses fake identifiers, as documented for journals acquired by so-called *journal snatchers* (Martín-Martín & Delgado López-Cózar, 2025), its output disappears from the database. This raises serious doubts about the rigor of these journals' editorial management after their removal from Scopus.

This decline in output is associated with removal from Scopus. If publishing in recognized journals brings prestige and supports a researcher's career progression and promotion, publishing in delisted journals may signal exactly the opposite, as an indication of research malpractice or unethical conduct, with the consequent loss of prestige or reputation. Indeed, two recent frameworks designed to measure the ethics and integrity of institutions, SCImago IRIS (Sánchez-Jiménez et al., 2026) and $RI^2$ (Meho, 2025), identify publication in delisted journals as a marker of questionable practices. Under these conditions, researchers have little incentive to publish in journals that the scientific community itself has flagged, which leads to fewer submissions to these journals and, in turn, to fewer publications. It should be remembered, however, that some evaluation systems do not penalize publication in delisted journals (Nazarovets, 2022), which could explain why a considerable number of these journals remain active.

Choosing a journal in which to publish is a key step in a researcher's career. It is a fully deliberate act. Citing a paper is a different matter. While choosing the *wrong* journal can harm a scientific career, citing the *wrong* paper rarely has consequences for the citing author. Researchers are unlikely to check the indexing level or status of the journal in which a cited article appeared (Lancho Barrantes et al., 2023). Our data reflect this. Although the pooled median of citations per document falls from 1.2 to 0.9, comparing each journal with itself shows hardly any change (from 0.93 to 0.90), and the difference is not statistically significant. In fact, almost half of the journals with comparable data showed no decline in the citation of their papers after delisting.

These results do not follow a uniform pattern across geographic regions or fields of knowledge. Journals from Northern America and Middle East are penalized in terms of citations after delisting, whereas among European journals those that maintain or increase their impact outnumber those whose impact declines. Likewise, in fields such as Engineering and Technology and Agricultural Sciences, more journals maintain or increase their impact after delisting than see it fall.

In other words, authors *punish* delisted journals by not submitting their work to them, but they do not penalize them to the same extent in terms of citations and continue to cite them at a similar rate. This is consistent with Lancho Barrantes et al. (2023), who found that journals removed from DOAJ for questionable practices continued to receive citations after removal, including from papers in high citation percentiles and from internationally co-authored work. It is also partly in line with the findings of Cortegiani et al. (2020), who, using a methodology based only on journal data prior to delisting, found that these papers continued to be cited, and even more than before their removal from the databases. Although we did not detect a higher level of citation, our results show that, among active journals, half maintain or increase the impact of the papers they publish after delisting. These data are a concern from a research integrity perspective, since work that has not passed proper academic scrutiny may be cited as established evidence.

Our study also shows that the journals removed from Scopus are concentrated in the lower-impact quartiles: 73.7% were in Q3 or Q4 at the time of delisting, compared with only 5.4% in Q1. The lower quartiles reproduce the general pattern of the study. Output falls significantly, while impact per document is barely affected, since among journals with citation data in both periods those that maintain or increase their C/D outnumber those whose C/D declines (51% in Q3 and 57% in Q4). Q1 journals show the opposite pattern. They are the only quartile in which the change in output is not statistically significant and, at the same time, the only one in which the change in citations per document is.

**Conclusions**

Our study has several implications for the academic community and for the governance of science and technology systems. The first concerns the central role of the major indexes as the *whitelists* of the academic world. Removal from a database such as Scopus can mean the academic *death* of a journal, either making it disappear entirely or relegating it to a marginal position in the academic publishing circuit. This gatekeeping role should not rest solely with large corporations. Evaluation agencies and science policy makers around the world need to take a more active role in how these journals are considered.

Second, our results suggest that authors are less careful when citing journals than when choosing where to publish. While authors appear to avoid publishing their articles in delisted journals, they seem to have no such reservations about citing them. This points to the need for more training and support for researchers, both in choosing journals and in citing, which should not be left to the algorithms of electronic platforms without quality filters or to the discretion of artificial intelligence tools.

Third, further studies are needed to analyze the effects and implications of delisted journals. Future research could, for example, examine the composition of the academic communities that use these journals after delisting and whether these are the same communities that already published in them. There is already evidence of changes in the research communities of journals that undergo a sudden loss of quality after a change in editorial management (Martín-Martín & Delgado López-Cózar, 2025), and it would be worth testing these findings on a larger scale. It would also be relevant to analyze how these journals are cited outside academia, particularly in policy documents, to determine the extent to which they inform decision-making in different spheres of society.

*Limitations*

One limitation of this study is the small number of indicators used to measure the journals' output and citation impact. Extending the analysis to other bibliometric and formal indicators would show whether journal quality also deteriorates from these other perspectives and would offer a broader and more nuanced picture of delisted journals.

Another limitation is that the length of the post-delisting period differs across cohorts. Journals delisted in 2023 and 2024 have only two years and one year of subsequent observation, respectively. Since the decline in output becomes most evident from the second year onwards, including these cohorts tends to attenuate the decline. If these cohorts were excluded, the decline in median output would increase from 37.1% to 43.3%. The results should therefore be read as a conservative estimate. In addition, the citation analysis is limited to papers published up to 2023, which reduces the number of documents and journals in the years after delisting.

Finally, the analysis relies entirely on OpenAlex, whose coverage depends heavily on the metadata that publishers deposit with Crossref. We identified 39 journals that were still active but whose later output no longer appeared in this index. Moreover, the study has no control group, so the changes associated with delisting cannot be separated from general trends in output and citation. Future studies could include such a control group to estimate the effect of delisting on journal output and citation impact more robustly.

**Declaration of Competing Interest**

This research did not receive any specific grant from funding agencies in the public, commercial, or not-for-profit sectors. The authors have no competing interests to declare that are relevant to the content of this article.

**CREDIT Taxonomy**

Álvaro Cabezas-Clavijo: Conceptualization, Data curation, Formal analysis, Investigation, Methodology, Visualization, Writing – original draft, Writing – review & editing.

Fernando Sánchez-Pita: Data curation, Formal analysis, Investigation, Writing – original draft, Writing – review & editing.

☐ Conception ☐ Execution ☐ Management ☐ Leadership ☐ Output

| AUTHORS | Conceptualization | Data curation | Formal analysis | Funding acquisition | Investigation | Methodology | Project administrati | Resources | Software | Supervision | Validation | Visualization | Writing – original dr | Writing – review & e | ROLES |
|---|---|---|---|---|---|---|---|---|---|---|---|---|---|---|---|
| ÁC Álvaro Cabezas-Clavijo #1 1st author | ✓ | ✓ | ✓ | | ✓ | ✓ | | | | | | ✓ | ✓ | ✓ | 8 |
| FS Fernando Sánchez-Pita #2 Last author | | ✓ | ✓ | | ✓ | | | | | | | | ✓ | ✓ | 5 |

*Generated with Credit Taxonomy Creator* [*https://acabezasclavijo.github.io/Credit/*](https://acabezasclavijo.github.io/Credit/)

**Disclosure of delegation to Generative AI**

The authors declare the use of generative AI in the research and writing process. According to the GAIDeT taxonomy (2026), the following tasks were delegated to GAI tools under full human supervision:

- Code generation (scripts for the automated download of data from OpenAlex)

- Proofreading and editing (review of inconsistencies and typos in the text; review of claims not supported by the data)

- Translation (of the original text, written in Spanish, into English)

The GAI tool used was: Claude (Sonnet 4.6 and Opus 5.5).

Responsibility for the final manuscript lies entirely with the authors.

GAI tools are not listed as authors and do not bear responsibility for the final outcomes.

*Generated with GAIDeT Declaration Generator* [*https://panbibliotekar.github.io/gaidet-declaration/*](https://panbibliotekar.github.io/gaidet-declaration/)

## References

Anderson, R. (2019). Citation Contamination: References to Predatory Journals in the Legitimate Scientific Literature. *The Scholarly Kitchen*. https://scholarlykitchen.sspnet.org/2019/10/28/citation-contamination-references-to-predatory-journals-in-the-legitimate-scientific-literature/

Bagues, M., Sylos-Labini, M., & Zinovyeva, N. (2019). A walk on the wild side: 'Predatory' journals and information asymmetries in scientific evaluations. *Research Policy, 48*(2), 462–477. https://doi.org/10.1016/j.respol.2018.04.013

Beall, J. (2012). Predatory publishers are corrupting open access. *Nature*, *489*(7415), 179–179. https://doi.org/10.1038/489179a

Butler, L. (2003). Explaining Australia's increased share of ISI publications—The effects of a funding formula based on publication counts. *Research Policy*, *32*(1), 143–155. https://doi.org/10.1016/S0048-7333(02)00007-0

Castillo-Nuñez, J. P. S., Minchon-Medina, C. A., Clemente-Vega, A., Vallenas-Aroni, N. R., Lozano-Lozano, M., & Báez-Sepúlveda, M. (2026). Editorial Predictors of the Discontinuation of Open Access Scientific Journals in Scopus: An Analysis from DOAJ. *Publications*, *14*(1), 2. https://doi.org/10.3390/publications14010002

Céspedes, L., Kozlowski, D., Pradier, C., Sainte-Marie, M. H., Shokida, N. S., Benz, P., Poitras, C., Ninkov, A. B., Ebrahimy, S., Ayeni, P., Filali, S., Li, B., & Larivière, V. (2025). Evaluating the linguistic coverage of OpenAlex: An assessment of metadata accuracy and completeness. *Journal of the Association for Information Science and Technology, 76*(6), 884–895. https://doi.org/10.1002/asi.24979

Chavarro, D., Alperin, J. P., & Willinsky, J. (2025). On the open road to universal indexing: OpenAlex and Open Journal Systems. *Quantitative Science Studies, 6*, 1039–1058. https://doi.org/10.1162/QSS.a.17

Chavarro, D., Tang, P., & Ràfols, I. (2017). Why researchers publish in non-mainstream journals: Training, knowledge bridging, and gap filling. *Research Policy, 46*(9), 1666–1680. https://doi.org/10.1016/j.respol.2017.08.002

Cobey, K. D., Lalu, M. M., Skidmore, B., Ahmadzai, N., Grudniewicz, A., & Moher, D. (2018). What is a predatory journal? A scoping review. *F1000Research*, *7*, 1001. https://doi.org/10.12688/f1000research.15256.2

Cortegiani, A., Ippolito, M., Ingoglia, G., Manca, A., Cugusi, L., Severin, A., Strinzel, M., Panzarella, V., Campisi, G., Manoj, L., Gregoretti, C., Einav, S., Moher, D., & Giarratano, A. (2020). Citations and metrics of journals discontinued from Scopus for publication concerns: The GhoS(t)copus Project. *F1000Research*, *9*, 415. https://doi.org/10.12688/f1000research.23847.2

Culbert, J. H., Hobert, A., Jahn, N., Haupka, N., Schmidt, M., Donner, P., & Mayr, P. (2025). Reference coverage analysis of OpenAlex compared to Web of Science and Scopus. *Scientometrics*, *130*(4), 2475–2492. https://doi.org/10.1007/s11192-025-05293-3

Dadkhah, M., Maliszewski, T., & Teixeira Da Silva, J. A. (2016). Hijacked journals, hijacked web-sites, journal phishing, misleading metrics, and predatory publishing: Actual and potential threats to academic integrity and publishing ethics. *Forensic Science, Medicine, and Pathology*, *12*(3), 353–362. https://doi.org/10.1007/s12024-016-9785-x

Demir, S. B. (2018). Predatory journals: Who publishes in them and why? *Journal of Informetrics, 12*(4), 1296–1311. https://doi.org/10.1016/j.joi.2018.10.008

Elmore, S. A., & Weston, E. H. (2020). Predatory Journals: What They Are and How to Avoid Them. *Toxicologic Pathology*, *48*(4), 607–610. https://doi.org/10.1177/0192623320920209

Elsevier. (2025). Scopus content policy and selection. Elsevier. https://www.elsevier.com/products/scopus/content/content-policy-and-selection

Forchino, M. V., & Torres-Salinas, D. (2026). The OpenAlex database in review: Evaluating its applications, capabilities, and limitations. *Journal of Informetrics, 20*(2), 101800. https://doi.org/10.1016/j.joi.2026.101800

Grudniewicz, A., Moher, D., Cobey, K. D., Bryson, G. L., Cukier, S., Allen, K., Ardern, C., Balcom, L., Barros, T., Berger, M., Ciro, J. B., Cugusi, L., Donaldson, M. R., Egger, M., Graham, I. D., Hodgkinson, M., Khan, K. M., Mabizela, M., Manca, A., … Lalu, M. M. (2019). Predatory journals: No definition, no defence. *Nature*, *576*(7786), 210–212. https://doi.org/10.1038/d41586-019-03759-y

Hicks, D., Wouters, P., Waltman, L., De Rijcke, S., & Rafols, I. (2015). Bibliometrics: The Leiden Manifesto for research metrics. *Nature*, *520*(7548), 429–431. https://doi.org/10.1038/520429a

Jamali, H. R., Wakeling, S., & Abbasi, A. (2022). Why do journals discontinue? A study of Australian ceased journals. *Learned Publishing*, *35*(2), 219–228. https://doi.org/10.1002/leap.1448

Jiménez-Contreras, E., De Moya Anegón, F., & Delgado López-Cózar, E. (2003). The evolution of research activity in Spain. *Research Policy*, *32*(1), 123–142. https://doi.org/10.1016/S0048-7333(02)00008-2

Krauskopf, E. (2018). An analysis of discontinued journals by Scopus. *Scientometrics*, *116*(3), 1805–1815. https://doi.org/10.1007/s11192-018-2808-5

Lancho Barrantes, B. S., Dalton, S., & Andre, D. (2023). Bibliometrics Methods in Detecting Citations to Questionable Journals. *The Journal of Academic Librarianship*, *49*(4), 102749. https://doi.org/10.1016/j.acalib.2023.102749

Marina, T., & Sterligov, I. (2021). Prevalence of potentially predatory publishing in Scopus on the country level. *Scientometrics, 126*(6), 5019–5077. https://doi.org/10.1007/s11192-021-03899-x

Martín-Martín, A., & Delgado López-Cózar, E. (2025). Invasion of the journal snatchers: How indexed journals are falling into questionable hands. https://doi.org/10.5281/zenodo.14766415

Mathies, C., Kivistö, J., & Birnbaum, M. (2020). Following the money? Performance-based funding and the changing publication patterns of Finnish academics. *Higher Education*, *79*(1), 21-37. https://doi.org/10.1007/s10734-019-00394-4

Meho, L. I. (2025). Gaming the metrics: bibliometric anomalies in global university rankings and the research integrity risk index ($RI^2$). Scientometrics, 130, 6683-6726. https://doi.org/10.1007/s11192-025-05480-2

Nazarovets, S. (2022). Analysis of publications by authors of Ukrainian institutes in Scopus-delisted titles. *Learned Publishing, 35*(4), 499–515. https://doi.org/10.1002/leap.1464

OECD. (2015). Frascati Manual 2015: Guidelines for collecting and reporting data on research and experimental development. OECD Publishing. https://doi.org/10.1787/9789264239012-en

Oviedo-García, M. Á. (2021). Journal citation reports and the definition of a predatory journal: The case of the Multidisciplinary Digital Publishing Institute (MDPI). *Research Evaluation, 30*(3), 405–419. https://doi.org/10.1093/reseval/rvab020

Öztürk, O., & Taşkın, Z. (2024). How metric-based performance evaluation systems fuel the growth of questionable publications?. *Scientometrics, 129*(5), 2729-2748. https://doi.org/10.1007/s11192-024-04991-8

Pölönen, J., Guns, R., Kulczycki, E., Sivertsen, G., & Engels, T. C. E. (2021). National Lists of Scholarly Publication Channels: An Overview and Recommendations for Their Construction and Maintenance. *Journal of Data and Information Science*, *6*(1), 50-86. https://doi.org/10.2478/jdis-2021-0004

Priem, J., Piwowar, H., & Orr, R. (2022). OpenAlex: A fully-open index of scholarly works, authors, venues, institutions, and concepts [Preprint]. arXiv. https://doi.org/10.48550/arXiv.2205.01833

Quaderi, N. (2023, March 20). *Supporting Integrity of the Scholarly Record in Web of Science*. Clarivate. https://clarivate.com/academia-government/blog/supporting-integrity-of-the-scholarly-record-our-commitment-to-curation-and-selectivity-in-the-web-of-science/

Quan, W., Chen, B., & Shu, F. (2017). Publish or impoverish: An investigation of the monetary reward system of science in China (1999-2016). *Aslib Journal of Information Management*, *69*(5), 486-502. https://doi.org/10.1108/AJIM-01-2017-0014

R Core Team. (2024). *R: A language and environment for statistical computing*. R Foundation for Statistical Computing. https://www.R-project.org/

Repiso-Caballero, R., & Delgado-Vázquez, Á.-M. (2024). Fallen Journals 2023. Implicaciones para la ciencia española de la expulsión de revistas en Web of Science. *Revista Mediterránea de Comunicación*, *15*(1), 373–384. https://doi.org/10.14198/MEDCOM.25211

Richardson, R. A. K., Hong, S. S., Byrne, J. A., Stoeger, T., & Amaral, L. A. N. (2025). The entities enabling scientific fraud at scale are large, resilient, and growing rapidly. Proceedings of the National Academy of Sciences, 122(32), e2420092122. https://doi.org/10.1073/pnas.2420092122

Saarela, M., Pölönen, J., Linna, A.-K., Wahlfors, L., Correia, A., & Kärkkäinen, T. (2026). Comparing LLM and expert assessments of journal quality. *Scientometrics*, *131*(5), 3525–3546. https://doi.org/10.1007/s11192-026-05644-8

Sánchez-Jiménez, R., Guerrero-Bote, V. P., Halevi, G., De-Moya-Anegón, F., & González, D. G. (2026). From concept to measurement: Operationalizing and normalizing Integrity Risk Indicators by SCImago (IRIS). *Research Integrity and Peer Review*, *11*(1), 48. https://doi.org/10.1186/s41073-026-00233-y

Siler, K. (2020). Demarcating spectrums of predatory publishing: Economic and institutional sources of academic legitimacy. *Journal of the Association for Information Science and Technology*, *71*(11), 1386–1401. https://doi.org/10.1002/asi.24339

Simard, M.-A., Basson, I., Hare, M., Larivière, V., & Mongeon, P. (2024). The value of a diamond: Understanding global coverage of diamond open access journals in Web of Science, Scopus, and OpenAlex to support an open future. Proceedings of the Annual Conference of CAIS / Actes du congrès annuel de l'ACSI. https://doi.org/10.29173/cais1845

Singh, V. K., Singh, P., Karmakar, M., Leta, J., & Mayr, P. (2021). The journal coverage of Web of Science, Scopus and Dimensions: A comparative analysis. *Scientometrics, 126*(6), 5113–5142. https://doi.org/10.1007/s11192-021-03948-5

Thelwall, M., & Jiang, X. (2025). Is OpenAlex suitable for research quality evaluation and which citation indicator is best? Journal of the Association for Information Science and Technology, 76(12), 1660–1681. https://doi.org/10.1002/asi.70020

Trueblood, J. S., Allison, D. B., Field, S. M., Fishbach, A., Gaillard, S. D. M., Gigerenzer, G., Holmes, W. R., Lewandowsky, S., Matzke, D., Murphy, M. C., Musslick, S., Popov, V., Roskies, A. L., Ter Schure, J., & Teodorescu, A. R. (2025). The misalignment of incentives in academic publishing and implications for journal reform. *Proceedings of the National Academy of Sciences*, *122*(5), e2401231121. https://doi.org/10.1073/pnas.2401231121

Virtanen, P., Gommers, R., Oliphant, T. E., Haberland, M., Reddy, T., Cournapeau, D., Burovski, E., Peterson, P., Weckesser, W., Bright, J., Van Der Walt, S. J., Brett, M., Wilson, J., Millman, K. J., Mayorov, N., Nelson, A. R. J., Jones, E., Kern, R., Larson, E., … Vázquez-Baeza, Y. (2020). SciPy 1.0: Fundamental algorithms for scientific computing in Python. *Nature Methods*, *17*(3), 261-272. https://doi.org/10.1038/s41592-019-0686-2

Wickham, H. (2016). *ggplot2: Elegant graphics for data analysis*. Springer. https://doi.org/10.1007/978-3-319-24277-4